\documentclass[a4paper]{jpconf}
\usepackage{graphicx}
\usepackage{natbib}
\usepackage{citesort}
\newlength{\defbaselineskip}
\newcommand{\setlinespacing}[1]%
           {\setlength{\baselineskip}{#1 \defbaselineskip}}

\begin{document}
\title{ The Nucleosynthesis and Reaction Rates of Fluorine 19 ($^{19}F$) in the Sun}

\author[0000-0001-9178-3992]{ Mohammad K.\ Mardini$^{1,2}$, Nidal Ershiadat$^{3}$, Mashhoor A. Al-Wardat$^{4}$, Ali A. Taani$^{5}$,Sergen {\"O}zdemir$^{6}$ Hamid Al-Naimiy$^{7}$ and  Awni Khasawneh$^{8}$}

\address{$^{1}$Key Lab of Optical Astronomy, National Astronomical Observatories, Chinese Academy of Sciences, A20 Datun Road, Chaoyang, Beijing 100102, People's Republic of China\\
               $^{2}$School of Astronomy and Space Science, University of Chinese Academy of Sciences, No.19(A) Yuquan Road, Shijingshan District, Beijing, 100049, People's Republic of China\\
               $^{3}$Department of Physics, University of Jordan, Amman, 11942 Jordan\\
               $^{4}$Department of Physics and Institute of Astronomy and Space Sciences, Al Al-Bayt University, Mafraq, 25113 Jordan\\
               $^{5}$Physics Department, Faculty of Science, Al-Balqa Applied University, 19117 Salt, Jordan\\
               $^{6}$Astronomy and Space Sciences Department, Ege University, Izmir, Turkey\\
               $^{6}$Department of Applied Physics and Astronomy, Sharjah University, Sharjah, United Arab Emirates\\
               $^{7}$Royal Jordanian Geographic Center, Amman, 11941 Jordan}
\ead{mardini@nao.cas.cn}

\begin{abstract}
We investigate the abundance of $^{19}F$ in the Sun through the nucleosynthesis scenario. In addition, we calculate the rate equations and reaction rates of the nucleosynthesis of $^{19}F$ at different temperature scale. Other important functions of this nucleosynthesis (nuclear partition function and statistical equilibrium conditions) are also obtained. The resulting stability of $^{19}F$ occurs at nucleus with A = 19 and Mass Excess= -1.4874 MeV. As a result, this will tend to a series of neutron captures and beta-decay until $^{19}F$ is produced. The reaction rate of $^{15}N$ ($\alpha$, $\gamma$) $^{19}F$ was dominated by the contribution of three low-energy resonances, which enhanced the final $^{19}F$ abundance in the envelope.
\end{abstract}

\section{Introduction}
The Fluorine abundances is playing a crucial role in the hotly debated nucleosynthesis scenarios in stellar populations. However, the Fluorine ($^{19}F$) is considered to be the only stable isotope of the fluorine element in the Sun. In addition, it is very sensitive to the physical conditions within stars in general, where the least abundant of stable nuclides in the 12-32 atomic mass range \citep{2009ARAA..47..481A}. Despite its importance, a detailed understanding of observational phenomena of fluorine nucleosynthesis (which can occur through the $^{15}N$ ($\alpha$, $\gamma$) $^{19}F$ reaction  is still under development and, the reaction rate still retains with large uncertainties due to the lack of experimental studies available. Several possible explanations were offered by many researchers to explain its production \citep{2011ApJ...729...40L}.

There are several astronomical cites that contain extremely hot gases, in order to let the $^{19}F$ production be detected there, such as supernovae Type II  \citep{1988Natur.334...45W, 2007PhRvC..76f5804L}, Wolf–Rayet stars \citep{2000A&A...355..176M}, and within the interiors of Asymptotic Giant Branch (AGB) stars during helium flashes \citep{1992A&A...261..164J, 2009ApJ...696..797C}.

However, available data from the Joint Institute for Nuclear Astrophysics (JINA) provides databases of nuclear data and reaction rates. We study the nuclear masses and nuclear partition functions to compute nuclear statistical equilibria which is a free
energy minimum state with some number of constraints, and then to compute the reaction rates of $^{19}F$.  The aim of this work is to study the process responsible for fluorine nucleosynthesis and their reaction rates, which may help us into understanding its origin in the Sun via H- and He-burning sequences. A quantitative comparison of observations with this process will lead the results from more detailed chemical evolution models incorporating the yields from very hot stellar interior.

Deriving the light element fluorine abundances are focused on using the HF line at 23358 {\AA}\, \citep[][and references therein]{2014ApJ...789L..41J}.
At this crowded/blended (CO lines) region, the detection of $^{19}F$ becomes harder.
Moreover, $^{19}F$  has very fragile nuclide, which can be easily destroyed by the most abundant species
in stellar interiors (proton or $\alpha$). Thus, the production site(s) of this element has been widely debated in literature \citep{1982ApJ...263..891W, 1996NuPhA.597..231D, 2004MNRAS.354..575R, 2005A&A...433..641W, 2005ApJ...619..884F, 2009ApJ...694..971A, 2011ApJ...739L..57K}

Jorissen et al \cite{1992A&A...261..164J} reported the leading work in this field, by presenting the first $^{19}F$ abundances of 49 K giants.
A new scenario for $^{19}F$ production after $\sim$ 1 Gyr was suggested by Renda et al \citep{2004MNRAS.354..575R}, this model treated
AGB stars as the major contributors of $^{19}F$ production; moreover SNe II considered as the only  contributor to the  $^{19}F$ abundance
in the early universe. To solve  the discrepancy between the theoretical and observed $^{19}F$ abundance in Galactic giants, recently, Kobayashi et al \citep{2011ApJ...739L..57K} adopts the yields of AGB stars and the $\nu$-process yields. This model expects
fluorine-to-iron plateau at low metallicities, relying on the neutrino luminosities. However, considering  $\nu$-process yields enhanced
$^{19}F$ production, and served us to fit the high observed abundances in very hot stellar populations.

\section{NUCLEOSYNTHESIS OF $^{19}F$}
The synthesis of $^{19}F$ can be produced  by consisting of a CNO H-burning sequence. A series of proton captures and beta-decays,
initiated on $^{14}N$, leads to a finite abundance of $^{19}F$, with $^{19}F(p,\alpha)^{16}O$ being the destructive
reaction through the reaction chain\\


$^{14}N(p,\gamma)^{15}O(\beta^{+})^{15}N(p,\gamma)^{16}O(p,\gamma)^{17}F$,
$^{17}F(\beta^{+})^{17}O(p,\gamma)^{18}F(\beta^{+})^{18}O(p,\gamma)^{19}F(p,\alpha)^{16}O$. \\
\\
The The adopted $^{19}F(p,\alpha)^{16}O$ rate is the
geometrical mean of the lower and upper limits to that rate
proposed by \citep{1990PhRvC..41..458R} Fluorine can also be
produced and destroyed during He-burning through the chains
\citep{1993nuco.conf..503M}.\\

~~~~~~~~~~~~~~~~~~~~~~~~~~~~~~$(\beta^{+})^{18}O(p,\alpha)^{15}N(\alpha,\gamma)^{19}F$

~~~~~~~~~~~~~~~~~~~~~~$\nearrow$~~~~~~~~~~~~~~~~~~~~~~~~~~~~~~~~~~~~~~~~~$\searrow$

$^{14}N(\alpha,\gamma)^{18}F$ ~~~ $\rightarrow$
~~~~~~~~~$(n,p)^{18}O(p,\alpha)^{15}N(\alpha,\gamma)^{19}F(\alpha,p)^{22}Ne$

~~~~~~~~~~~~~~~~~~~~~~$\searrow$~~~~~~~~~~~~~~~~~~~~~~~~~~~~~~~~~~~~~~~~~$\nearrow$

~~~~~~~~~~~~~~~~~~~~~~~~~~~~~~~~~~~~~~$(n,\alpha)^{15}N(\alpha,\gamma)^{19}F$\\
\\
As a result, the synthesis of $^{19}F$ thus requires the availability of
neutrons and protons. They are mainly produced by the reactions
$^{13}C(\alpha,n)^{16}O$ and $^{14}N(n,p)^{14}C$.

\section{Computational Tool}

We used nucnet tools to explore the abundance of $^{19}F$ in the Sun. This code was
built by the Astronomy and Astrophysics group at
Clemson University, South Carolina, USA \footnote{https://sourceforge.net/p/nucnet-tools/home/Home/}.

The Joint Institute for Nuclear Astrophysics (JINA) provides databases of nuclear data and reaction rates.
We were focused on the Snapshots, in order to study the nuclear masses and nuclear partition functions.
It was  also  used to compute nuclear statistical equilibria and finally to compute the reaction rates of  $^{19}F$.

\section{Results and discussion}

\subsection{Nuclear Masses}

\begin{figure}[!ht]
\centering
\includegraphics[width=0.99\textwidth,keepaspectratio]{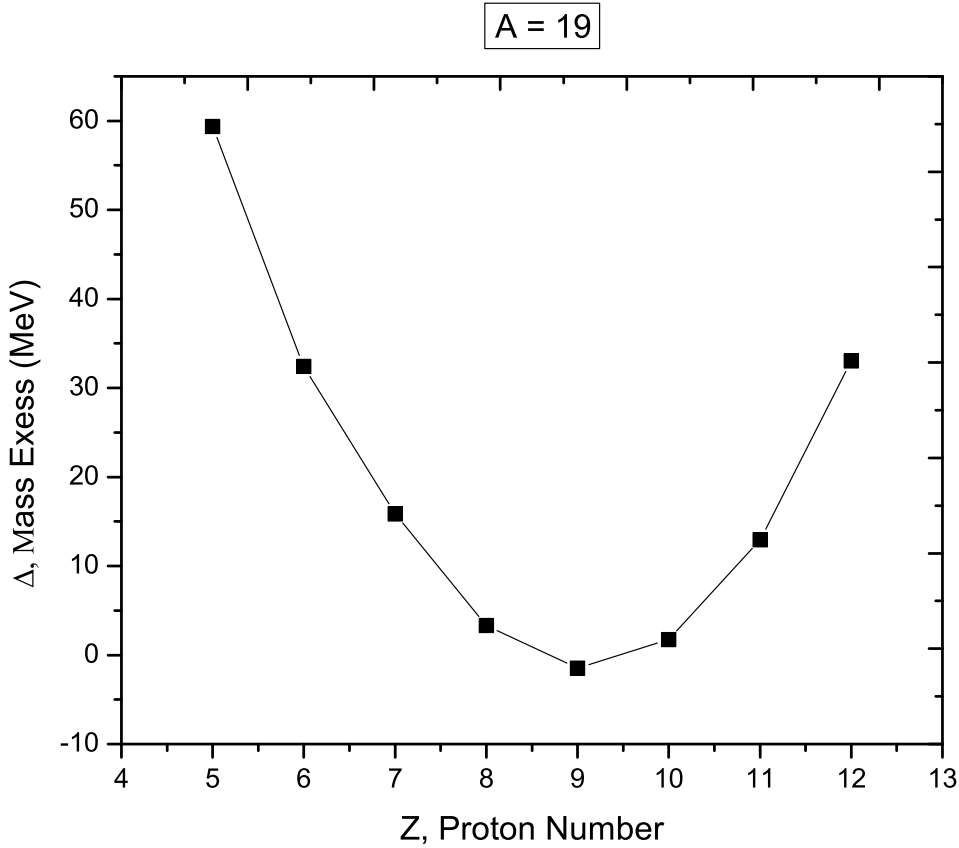}
\caption{Beta valley stability for $A = 19$ nuclides}\label{A19excess}
\end{figure}

Here, the stability of  $^{19}F$ can be studied by the mass excess (See figure \ref{A19excess}),
the expected parabola shape of the mass distribution for a given set of isobars
suggest that nucleus with A = 19 will tend to beta decay until the $^{19}F$ produced.

Our calculation shows that if the nucleus (A = 19) has $Z <$ 9, it will
turn some of its neutrons into protons by $\beta^{-}$ decay
until $Z =$ 9. On the other hand, if $Z >$ 9 it will turn
protons into neutrons by $\beta^{+}$ decay until $Z =$ 9.

\begin{table}
  \centering
  \begin{tabular}{lcccccc}
    \hline
    \hline
    Index & Z & A & Name &Mass Excess (MeV)& Spin &Data Source \\
    \hline
 0& 5&19& $^{19}B$& 59.3640& 3/2 &reac1  \\
 1& 6&19& $^{19}C$& 32.4207& 3/2 &reac1     \\
 2& 7&19& $^{19}N$& 15.8621& 1/2 &reac1    \\
 3& 8&19& $^{19}O$& 3.3349  & 5/2 &reac1    \\
 4& 9&19& $^{19}F$ & -1.4874 & 1/2 &reac1   \\
 5&10&19& $^{19}Ne$& 1.7514& 1/2 &reac1   \\
 6&11&19& $^{19}Na$& 12.9268& 3/2 &reac1   \\
 7&12&19& $^{19}Mg$& 33.0401& 1/2 &reac1   \\
    \hline
  \end{tabular}
  \caption{Nuclides with A= 19, the collection has 8 species.}\label{A19EX}
\end{table}

\subsection{The Nuclear Partition Function}

\begin{enumerate}
\item
We have considered the $^{13}C(\alpha,n)^{16}O$ reaction, by adopting the
rate from \citep{1993ApJ...414..735D} and
\citep{1995AIPC..327..255D}, which is about $50\%$ lower than
the rate recommended by \emph{NACRE} in the temperature range of
interest.

\item
The reaction $^{14}C(\alpha,\gamma)^{18}O$ has been studied
experimentally in the energy range of 1.13-2.33 MeV near the
neutron threshold in the compound nucleus $^{18}O$ by
\citep{1992NuPhA.548..414G}. The reaction rate is dominated at
higher temperatures by the direct capture and the single strong
$4^{+}$ resonance at center-of-mass energy $E_{cm}=0.89 MeV$ toward lower temperatures.
As a result, the contributions may come from the
$3^{-}$ resonance at $E_{cm}=0.176 MeV E_{x}=6.404 MeV$ and a
$1_{-}$ subthreshold state at $E_{x}=6.198 MeV$, which are important for He shell
burning in AGB stars. It has been
shown in detailed cluster model simulations that neither one of
the two levels is characterized by a pronounced  $\alpha$
cluster structure\citep{1985PhRvC..31.2274D}. The strengths of
these two contributions are unknown and have been estimated by
\citep{1988ApJ...324..953B}.

The $^{14}C(\alpha,\gamma)^{18}O$ reaction can be activated
together with $^{13}C(\alpha,n)^{16}O$ reaction during the
interpulse period. This could be happened in both the partial mixing zone and the
deepest layer of the region composed by H-burning ashes, when
$^{14}N(n,p)^{14}C$ occurs. This represents the main path to
the production of $^{18}O$ and subsequently of $^{15}N$. The
importance of the nucleosynthesis of $^{15}N$ during the
interpulse periods is very much governed by the choice of the
rate of the $^{14}C(\alpha,\gamma)^{18}O$ reaction. The closer,
or higher, this rate is to that of the $^{13}C(\alpha,n)^{16}O$
reaction, the more effcient is the production of $^{15}N$
because $^{18}O$ and protons are produced together. The effect
of the partial mixing zone and hence the uncertainties related
to it are in fact much less important when using our
recommended rate, since in the temperature range of interest
our rate is more than an order of magnitude lower than our
standard rate from \citep{2001NuPhA.688..508J}

\item
The low-energy resonances in $^{14}N(\alpha,\gamma)^{18}F$ have
recently successfully been measured by
\citep{2000PhRvC..62e5801G}. Previous uncertainties about the
strengths of these low-energy resonances were removed. Because
of these results, the reaction rate is reduced by about a
factor of 3 compared to \emph{NACRE}. The
$^{14}N(\alpha,\gamma)^{18}F$ reaction is inefficient at the
temperature of neutron release in the partial mixing zone while
it is activated in the convective pulse. Hence, its rate only
affects the production of $^{19}F$ in the pulse. Using the new
rate by \citep{2000PhRvC..62e5801G} with respect to the rate by
\citep{1988ADNDT..40..283C}

\item
The reaction rate of $^{15}N(\alpha,\gamma)^{19}F$ was taken
from \emph{NACRE}. The rate is dominated by the contribution of three
low-energy resonances. The resonance strengths are based on the
analysis \citep{1996NuPhA.597..231D}. It should be noted,
however, that there were several recent experimental studies
that point toward a significantly higher reaction rate.
\citep{1997PhRvC..55.3149D} already suggested higher resonance
strengths than given in their earlier paper. The final $^{19}F$
abundance in the envelope increased by a few percent only. This
is because the temperature in the thermal pulses is high enough
that in any case all $^{15}N$ is transformed in $^{19}F$ as
shown in Figure~\ref{Fluorineabandance}

\begin{figure}[!ht]
\centering
\includegraphics[width=0.99\textwidth,keepaspectratio]{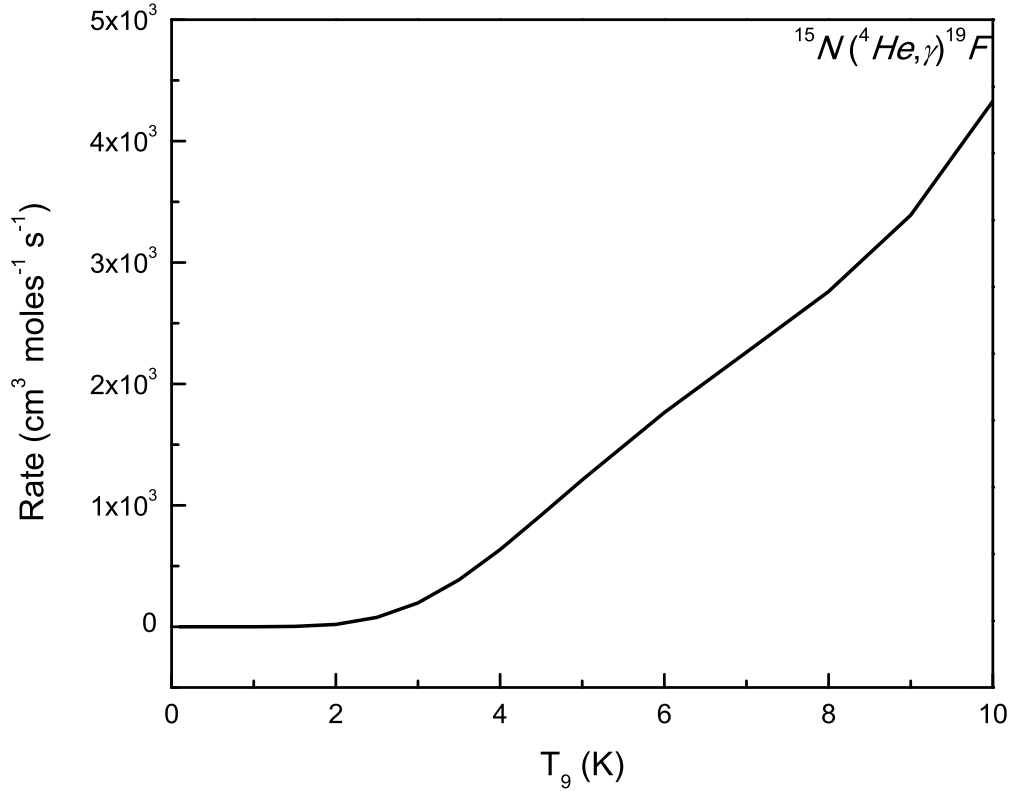}
\caption{Reaction rate of $^{15}N(^{4}He,\gamma)^{19}F$}\label{Fluorineabandance}
\end{figure}

\item
The $^{15}N(p,\alpha)^{14}C$ reaction has been investigated by
\citep{1952PhRv...86..527S}, and more recently by
\citep{1982ZPhyA.305..325R} at $E_{p}(lab) = 78-810 $~Kev.

\item
The $^{18}O(\alpha,\gamma)^{22}Ne$ reaction is of interest for
the discussion of $^{19}F$ production in AGB stars since it
competes with the $^{18}O(p,\alpha)^{15}N$ process. A strong
rate might lead to a reduction in $^{19}F$ production. The
reaction rate of $^{18}O(\alpha,\gamma)^{22}Ne$
\citep{1994ApJ...437..396K}. The main uncertainties result from
the possible contributions of low-energy resonances that have
been estimated on the basis of $\alpha$-transfer measurements
by \citep{1994NuPhA.567..146G}. A recent experimental study of
$^{18}O(\alpha,\gamma)^{22}Ne$ \citep{2003PhRvC..68b5801D} led
to the first successful direct measurement of the postulated
low-energy resonances at 470 and 566 keV.

\item
The reaction $^{18}O(p,\alpha)^{15}N$ provides a major link for
the production process of $^{19}F$. The reaction cross section
has been measured by \citep{1979NuPhA.313..346L} down to
energies of $\approx 70 $ KeV. Possible contributions of low
energy near threshold resonances were determined by
\citep{1982ApJ...263..891W} and \citep{1986NuPhA.457..367C}
using direct capture and single-particle transfer reaction
techniques.

\item
The Reaction Rate of $^{19}F(\alpha,p)^{22}Ne$ is one of the
most important input parameters for a reliable analysis of
$^{19}F$ nucleosynthesis at AGB stars. However, there is very
little experimental data available for the
$^{19}F(\alpha,p)^{22}Ne$ reaction cross section at low
energies. Experiments were limited to the higher energy range
above $E_{\alpha}= 1.3$ MeV \citep{1965Phy....31.1603K}. CF88
suggested a rate that is based on a simple barrier penetration
model previously used by \citep{1969ApJS...18..247W}. This
reaction rate is in reasonable agreement with more recent
Hauser-Feshbach estimates assuming a high level density see
also \citep{1986ana..work..525T} and has therefore been used in
most of the previous nucleosynthesis simulations (See Figure \ref{f19(he4,p)ne22}).
\end{enumerate}

\begin{figure}[!ht]
\centering
\includegraphics[width=0.99\textwidth,keepaspectratio]{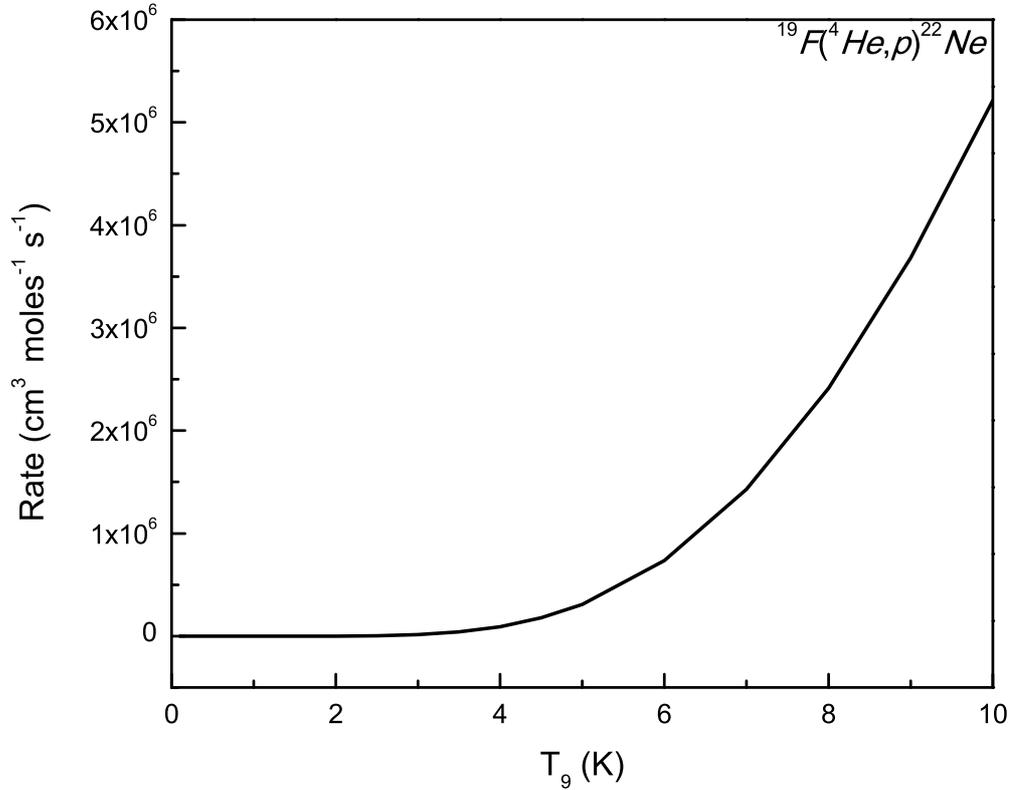}
\caption{Reaction rate of $^{19}F(^{4}He,p)^{22}Ne$}\label{f19(he4,p)ne22}
\end{figure}

References for $\alpha$ and neutron captures that we have used
in this paper are presented in Tables \ref{alpha} and \ref{Neutron}
respectively. All of the reactions not listed in the tables are
taken from the \emph{REACLIB} Data Tables.

\begin{table}
  \centering
  \caption{$\alpha$~-Capture}\label{alpha}
  \begin{tabular}{lcc}
    \hline
    \hline
    Reaction & References \\
    \hline
    $^{13}C(\alpha,n)^{16}O$~...............................& \citep{1993ApJ...414..735D}, \citep{1995AIPC..327..255D} \\
    $^{14}C(\alpha,\gamma)^{18}O$~...............................& \citep{2001NuPhA.688..508J} \\
    $^{15}N(\alpha,\gamma)^{19}F$~...............................& \citep{1996NuPhA.597..231D} \\
    $^{17}O(\alpha,n)^{20}Ne$~............................... & \citep{1995AIPC..327..255D} \\
    $^{18}O(\alpha,\gamma)^{22}Ne$~............................... & \citep{1994ApJ...437..396K}, \citep{1994NuPhA.567..146G} \\
    $^{18}O(\alpha.n)^{21}Ne$~............................... & \citep{1995AIPC..327..255D} \\
    \hline
  \end{tabular}
\end{table}

\begin{table}
  \centering
  \caption{Neutron Capture}\label{Neutron}
  \begin{tabular}{lcc}
    \hline
    \hline
    Reactions & References  \\
    \hline
    $^{12}C(n,\gamma)^{13}C$~...............................~~~~~~~~~~~~~~~~~~~~~~~~~~~~~~ & \citep{1998PhRvC..57.2724K} \\
    $^{13}C(n,\gamma)^{14}C$~...............................~~~~~~~~~~~~~~~~~~~~~~~~~~~~~~ & \citep{1990PhRvC..41..458R} \\
    $^{14}N(n,p)^{14}C$~...............................~~~~~~~~~~~~~~~~~~~~~~~~~~~~~~& \citep{1995AIPC..327..173G} \\
    $^{16}O(n,\gamma)^{17}O$~...............................~~~~~~~~~~~~~~~~~~~~~~~~~~~~~~ & \citep{1995ApJ...441L..89I} \\
    $^{18}O(n,\gamma)^{19}O$~...............................~~~~~~~~~~~~~~~~~~~~~~~~~~~~~~ & \citep{1996PhRvC..53..459M} \\
    \hline
  \end{tabular}
\end{table}

\section{Conclusion}
We studied the nuclear masses, nuclear statistical equilibria, and nuclear partition
functions of the $^{19}F$ in the Sun. In this work, we have concentrated more on the study
and interpretation of the evolution of the $^{19}F$ by utilizing the entire observational data of the solar system. Our results are in good agreement with other recent \emph{F} abundance determinations.

Our conclusions confirmed the stability of $^{19}F$, with atomic mass number, A= 19, atomic number,
\emph{Z} = 9 and Mass Excess, $\Delta$=  -1.4874 MeV.
This will lead to some characteristic of "valley of beta stability" (see Fig. 1).
 One can relate this stability  due to the $^{19}F$ has the minimum mass excess. This feature will be scaled by the same factor of $m_{U}c^{2} \emph{A}$.
 More detailed theoretical studies are required to explain this behavior.

During the course of this study, we found that the reaction rate of $^{15}N(\alpha,\gamma)^{19}F$ was dominated by the contribution of three
low-energy resonances, which enhanced the final $^{19}F$ abundance in the envelope. In addition, we found that $^{56}\emph{Ni}$ has the most of the mass during to nuclear statistical equilibrium, because the $^{56}\emph{Ni}$ species has the largest binding energy per nucleon.

It is noteworthy to mention here that, despite the lack of data, we limited the $^{19}F(\alpha,p)^{22}Ne$ reaction,  to higher energy range
above $E_{\alpha}= 1.3$ MeV, to achieve better agreement with Hauser-Feshbach  \citep{1986ana..work..525T}.

More details and results on nuclear statistical equilibrium and the expansion time are required to control the dynamical calculations
and characteristics.

\section*{Acknowledgements}
We are very grateful to the Astronomy and Astrophysics group at Clemson University, South Carolina, USA for using the nucnet tools.

 \bibliographystyle{iopart-num}
\bibliography{References}

\end{document}